# A global view of drug-therapy interactions

Jose C Nacher[1], Jean-Marc Schwartz[2]

[1]Department of Complex Systems, Future University-Hakodate, Hokkaido 041-8655, Japan.
[2]Faculty of Life Sciences, University of Manchester, Manchester M13 9PT, UK.

*Corresponding authors: JCN nacher@fun.ac.jp, JMS jean-marc.schwartz@manchester.ac.uk*

## Abstract

Background: Network science is already making an impact on the study of complex systems and offers a promising variety of tools to understand their formation and evolution in many disparate fields from technological networks to biological systems. Even though new high-throughput technologies have rapidly been generating large amounts of genomic data, drug design has not followed the same development, and it is still complicated and expensive to develop new single-target drugs. Nevertheless, recent approaches suggest that multi-target drug design combined with a network-dependent approach and large-scale systems-oriented strategies create a promising framework to combat complex multi-genetic disorders like cancer or diabetes.

Results: Here, we investigate the human network corresponding to the interactions between all US approved drugs and human therapies, defined by known drug-therapy relationships. Our results show that the key paths in this network are shorter than three steps, indicating that distant therapies are separated by a surprisingly low number of chemical compounds. We also identify a sub-network composed by drugs with high centrality measures, which represent the structural back-bone of the drug-therapy system and act as hubs routing information between distant parts of the network.

Conclusions: These findings provide for the first time a global map of the large-scale organization of all known drugs and associated therapies, bringing new insights on possible strategies for future drug development. Special attention should be given to drugs which combine the two properties of (a) having a high centrality value and (b) acting on multiple targets.

# Introduction

Complex behavior and networked structures emerge in systems composed of many interacting elements (1-4). Exploration of large databases from drastically different systems has allowed us to construct complex networks and uncover their organizing principles in many disparate fields, from large communication networks (5,6), transportation infrastructures (7) and social communities (8,9) to biological systems (1,10). Recent drug development strategies suggest that multi-target drug design combined with network-oriented approaches are promising to combat complex multi-genetic disorders (11-13). However, this raises the question of what are the direct and indirect network-dependent effects. Addressing such questions requires an accurate description and fundamental knowledge of the large-scale interactions between drugs and therapies/diseases. Breathtaking advances in pharmacology and medical science together with improvements in storage and data management have made it possible to organize and classify huge amount of information from drugs and associated diseases and therapies. The DrugBank database is one of the largest chemo-informatics resources that contains detailed information about approved drugs and drug targets. The drug category also includes information about their associated therapeutic properties following the Anatomic Therapeutic Chemical (ATC) classification. This knowledge allows us for the first time to investigate the human network corresponding to the interaction between all US approved drugs and associated human therapies defined by known drug-therapy relationships. This network defines a bipartite graph (15) whose nodes can be classified into two disjoint sets of drugs (D) and therapies (T) such that each edge connects a node in D and one in T. Thus, it is not possible to find two adjacent nodes within the same set. This bipartite graph can be decomposed into two networks. The drug projection is composed of nodes from the set D and two drugs are connected if there is a common therapy that is involved in both. The therapy projection is based on nodes from the set T and two therapies are connected if a drug implicated in both therapies exists. Therapies are closely linked to diseases, therefore the therapy network gives insights about the relations between diseases as well and completes previous work about the global organization of the human disease network (16,17).

Here, network analyses based on the bipartite graph and the associated network projections have revealed striking properties that characterize this global map of drug-therapy interactions. Our findings indicate that the network has a small average shortest-path length. In particular, the average distance between therapies is less than three steps, suggesting that distant therapies are separated by a low number of chemicals. In addition, our results indicate that much of the chemical information flowing through the network is routed through a small number of drug hubs.

In order to identify the main set of drugs/therapies that governs the network, we computed several network centrality metrics (14) that characterize the most influential nodes in the network. Next, exploitation of the correlations between pairs of different metrics provides a complementary perspective on the heterogeneous statistical properties of the network. Then, we identified a sub-network composed by drugs with high *betweenness* centrality, which represents the structural back-bone of the drug-therapy system. Identified drugs with highest centrality include *Scopolamine, Morphine, Tretinoin* and *Magnesium Sulfate*. Special attention must be given to drugs which combine the two properties of (1) having a high centrality value and (2) acting on multiple targets.

# Results

## a) Drug and therapy networks

The hierarchical structure of the ATC classification makes it possible to represent drug and therapy networks at five different levels, progressively revealing more details on interaction patterns. A striking observation is that the therapy network is fully connected at level 1 (Figure 1), and still almost fully connected at level 2 with the exception of one small isolated component. This finding is unexpected, since many drugs only have one specific therapeutic application (see Fig. 2). By computing the number of connections of each therapy network node (i.e., node degree $k$) at level 3, we found that the degree distribution follows a power-law $P(k) \propto k^{-\gamma}$, with degree exponent $\gamma$ =1.1 (18). That is, the probability to find highly connected therapies, or hubs, is rather higher than in an equivalent random network. Furthermore, the smaller the value of $\gamma$, the more influential the role of the hubs is in the network. We thus conclude that these highly connected therapies play a relevant role in this network because the observed degree exponent is not high. In addition, scale-freeness of the therapy network is conserved in the hierarchy of the ATC classification, as it is observed in both levels 2 and 3 (see Fig. 3 a-b).

The full bipartite network (Supplementary Material Figs. 1-3) shows that a majority of drugs are grouped in clusters connected to a specific therapy. But links exist between therapies, which are created by drugs spanning different therapeutic classes. These drugs acquire a particular significance, since they create links between different therapies and allow the complete therapy network to be connected. This observation can be quantitatively examined by constructing the histogram of the number of complete therapy identifiers associated to each drug (corresponding to level 5 of the ATC classification) shown in Figure 2. This histogram reveals that a majority of drugs (79%) are associated to a unique therapy. These drugs create no connection in the therapy network projection, all edges are due to the remaining 21% of drugs. It is surprising that the therapy network remains fully connected despite this small proportion. Moreover, edges do not predominantly connect therapies belonging to the same first-level class. This finding is made visible in Figure 1, where nodes have been colored according to their first-level class, revealing a large number of links between distinct classes. It is worth noticing that if only 21% of drugs create connections at level 5, it implies that the proportion is even smaller at inferior levels, as smaller levels lead to a merging of therapy nodes.

## b) Shortest paths

Shortest paths provide a measure of the efficiency of information flux in a network. For example, the efficiency of the chemical mass flux in a metabolic network can be estimated by computing its average shortest path length. Here, by investigating the therapy network projection, constructed using the level 2 of the ATC hierarchical classification, we have found that the average distance between two randomly selected therapies is less than three steps (2.61), which is very low (see Fig. 1c). This network is composed of 66 nodes and 237 edges and the main connected component has 64 nodes and 236 edges. It implies that in average distant therapies are separated by a surprisingly low number of chemical compounds. This value slightly increases to 3.41 when level 3 of the therapy network is considered (see Suppl. Material

Fig. 4). This network has 123 nodes and 349 edges, and the main connected component consists of 106 nodes and 338 edges.

It is known that the average path length $<l>$ is smaller in the Barabási-Albert network than in a random graph for any network size $n$. It means that a scale-free topology performs better in connecting distant nodes than random structures. Bollobás and Riordan (19) have shown that the average path length of a scale-free model network follows $<l> \sim \ln(n)/\ln\ln(n)$. The computation of this expression for the therapy networks in levels 2 and 3 gives a value of 2.92 and 3.05, respectively. These values are compatible with the observed values of 2.61 and 3.41. This reflects the scale-free topology observed in therapy network (see Fig. 3 a-b)

**c) High-centrality drugs and network backbones**

We investigated the "*betweenness*" of network nodes, a graph theoretical centrality metric. While the degree $k$ of a node explains the general topological features of the network and can only capture the local structure of network nodes (nearest neighbors), the betweenness $B_i$ of a given node $i$ is related to how frequently a node occurs on the shortest paths between all the pairs of nodes in the network (14,20). Hence, betweenness centrality identifies nodes with great influence over how the information reaches distant network nodes. This metric is relevant because it connects the local network structure to the global network architecture. In other contexts, it has been proven to be an indicator of interdisciplinarity (21) and was successfully used in different research areas from yeast protein interactome, for detecting essential proteins and their evolutionary age (22), to the problem of epidemics, for identifying key players in spreading an infection (23).

In Table 1, we show the top-20 drugs with highest betweenness in the drug network projection corresponding to level 2 in the ATC hierarchical classification. This information is complemented with the measure of the *closeness* centrality $C_i$, which measures how close a given node $i$ is to others (24). In some contexts, closeness can be understood as a measure of how long it will take information to spread from a given node to distant nodes in the network. Thus, nodes with high closeness indicate that its influence can reach others more rapidly. In this network, closeness is relatively high and homogeneous for most nodes. Table 1 indicates that drugs with highest betweenness are correlated with relatively high values of closeness in most cases. The generic drug name and the associated therapy classes are also displayed. The main connected component of this drug network consists of 828 nodes with an average path length of 3.15.

We were able to identify a reduced drug-therapy bipartite network composed by drugs with highest betweenness centrality (see Fig. 4). This sub-network reflects the structural back-bone of the drug-therapy system and has great influence over the chemical information flowing through the network. Nodes with high $B_i$-centrality are relevant because they bridge interactions between distinct parts of the network. Identified drugs with highest centrality include *Scopolamine, Morphine, Tretinoin* and *Magnesium Sulfate*. It is worth noticing that the backbone is almost fully connected as well, with the exception of two smaller isolated components. Distant therapies can be connected by a few drugs with high betweenness. For example, *Tolbutamide* and *Magnesium Sulfate* define a key shortest path of distance two between distant therapies like "Insulines and analogues" (A10) and "Dermatological preparations"

(D11). "Cardiac therapy" (C01) is directly connected to the "Antihemorrhagics" node (B02) via the drug *Epinephrine*. Apparently unrelated disorders like diabetes and dermatological lesions are thus separated by a much lower number of chemicals than could be expected.

**d) Correlations between network measures**

We investigated the correlations between each of the three measures of topological importance (degree $k$, betweenness $B_i$ and closeness $C_i$) in the drug network by calculating their Pearson correlation coefficient and P-value. Our results show that node degree $k$ significantly correlates with closeness and betweenness centralities (see Fig. 3 c-d). This finding can be explained by similar mechanisms to these generating the scale-free property in other networks. In scale-free models, it is well-known that high-connectivity nodes also exhibit high betweenness and closeness centralities. Therefore, this finding is consistent with the observation of a scale-free topology revealed in the therapy network. However, it is worth noticing that betweenness centrality does not correlate well to the number of drug targets. The finding that significant correlations between these measures are present suggests the existence of organizing principles behind the man-made drug-therapies/disease system. The chemical information that connects distant diseases and drugs composed of chemical compounds is routed through few drug-therapy nodes having wide influence in the global network. Future multi-target drugs might be designed by using similar baselines.

**e) Multi-target drugs**

Previous studies have revealed that the distribution of targets associated to approved drugs follows a power-law (25). A majority of drugs act on only one target, but a small number of drugs act on a large number of targets, which can reach up to 14. The development of drugs that are able to affect multiple targets is seen as promising for treating complex diseases (26,27). A special role must be played by these drugs which combine the two properties of (1) having a high $B_i$-centrality value and (2) acting on multiple targets. These drugs occupy pivotal positions in the drug-therapy network, as they not only connect heterogeneous therapies but also influence multiple metabolic pathways. Several drugs identified in Table 1 meet these two criteria, including *Hydroxocobalamin, Vitamin B3, Vitamin B12, Atropine, Orphenadrine, Procaine*.

Even though new high-throughput technologies have rapidly been developed generating large amounts of genomic data, drug design has not followed the same development, and it is still complicated and expensive to develop new single-target drugs. In contrast, new approaches suggest that multi-target drugs not only maximize the number of possible points of action but also introduce novel network disruption and systems-oriented strategies (12,13). Therefore, multi-target drug design combined with a network-dependent approach (11) create a promising concept to combat diseases based on multi-genetic disorders such as cancer, and diseases that involve a variety of cell types such as immunoinflamatory disorders and diabetes (28). Although the application of these strategies to drug development is still incipient, early results are encouraging (29) and suggest that the control of a complex disease system should consist in the simultaneous disruption of multiple targets located in distant network pathways.

# Methods

**a) Database and ATC classification**

The DrugBank database is a bioinformatics and chemoinformatics resource developed by the University of Alberta that contains detailed drug and drug target information (30). The database contains nearly 4300 drug entries including 1200 small molecule drugs approved by the US Food and Drug Administration (FDA). Each entry contains more than 80 data fields with detailed chemical, pharmacological and pharmaceutical drug data as well as sequence, structure and pathway information of drug targets.

The Drug Category field contains information on the therapeutic properties and general category of the drug. Therapeutic properties are entered following the Anatomic Therapeutic Chemical (ATC) classification (http://www.whocc.no/atcddd/). The ATC system is used by the World Health Organization as an international standard for drug utilization studies. It divides drugs into different groups according to the organ or system on which they act and their chemical, pharmacological and therapeutic properties. Drugs are classified in groups at five different levels. The first level of the code is based on a letter for the anatomical group and consists of one letter; there are 14 main groups. The second level of the code is based on the therapeutic main group. The third and fourth levels are chemical/pharmacological/therapeutic subgroups and the fifth level is the chemical substance. The hierarchical structure of the ATC classification provides an ideal framework for analyzing the relations between drugs and therapeutic applications. Each level of the ATC classification reveals complementary information, making it possible to navigate between different resolutions.

**b) Network construction**

We extracted the set of ATC identifiers associated to each FDA-approved drug to construct a bipartite network of drugs and therapies. 186 drugs had no ATC identifier and were discarded, leaving a network of 1014 drugs. Two projections of this bipartite network can be constructed. In the drug projection, nodes represent drugs and two nodes are connected if they share a common therapeutic property. In the therapy projection, nodes represent therapies and two nodes are connected if a common drug belongs to the two therapeutic category. Thanks to the hierarchical structure of the ATC classification, five possible networks can be constructed for each of these two projections. In the therapy projection, the number of nodes decreases at smaller ATC levels as several therapies are merged. In the drug projection, the number of nodes is independent of the ATC level but the number of edges increases at smaller levels.

**c) Definition of network metrics**

**Betweenness** ($B_i$) is a centrality measure of a node in a network. This metric goes beyond local information and reflects the role played by a node in the global network architecture. It is calculated as the fraction of shortest paths between node pairs that pass through a given node.

For a graph $G:=(V,E)$ with $n$ nodes, the betweenness for a node $i$ reads as:

$$B_i = \frac{1}{(n-1)(n-2)} \sum_{s \neq i \neq t \in V} \frac{\sigma_{st}(i)}{\sigma_{st}}$$

where $\sigma_{st}$ is the number of shortest paths from $s$ to $t$, and $\sigma_{st}(i)$ the number of shortest paths from $s$ to $t$ that pass through a node $i$ (20). This measure is normalized by the number of pairs of nodes without including $i$, that is $(n-1)(n-2)$.

**Closeness centrality** ($Ci$) measures how close a node $i$ is to all others in the same network and is defined as the average mean path between a node $i$ and all other nodes reachable from it:

$$C_i = \frac{n}{\sum_{j \in V} d(i,j)}$$

where $d(i,j)$ is the shortest distance between nodes $i$ and $j$, and $n$ is the number of nodes in the network (24).

**Average shortest path** is defined as the average number of steps along the shortest paths for all possible pairs of network nodes. This metric indicates the efficiency of information flux on a network.

**Node degree** $k$ is the number of edges connected to a given node.

# Tables

**Table 1:**

Top-20 drugs with highest betweenness in the drug network projection corresponding to level 2 in the ATC hierarchical classification. The associated therapy classes as well as closeness centrality, node degree and number of targets are also displayed.

| Accession No | Generic name | Therapy classes | B centrality | C centrality | Degree | Number of targets |
|---|---|---|---|---|---|---|
| APRD00616 | Scopolamine | A04, N05, S01 | 0.0841 | 0.469 | 155 | 1 |
| APRD00215 | Morphine | G04, N02, N04, R05, S01 | 0.0549 | 0.479 | 164 | 1 |
| APRD00362 | Tretinoin | D10, L01 | 0.0420 | 0.365 | 80 | 3 |
| APRD01080 | Magnesium Sulfate | A06, A12, B05, D11, V04 | 0.0412 | 0.361 | 24 | 3 |
| APRD00373 | Celecoxib | L01, M01 | 0.0340 | 0.356 | 94 | 1 |
| APRD00047 | R-mephobarbital | N03, N05 | 0.0294 | 0.336 | 81 | 1 |
| APRD00406 | Physostigmine | S01, V03 | 0.0261 | 0.454 | 106 | 1 |
| APRD00807 | Atropine | A03, N04, S01 | 0.0251 | 0.460 | 126 | 5 |
| APRD00479 | Lidocaine | C01, C05, D04, N01, R02, S01, S02 | 0.0229 | 0.470 | 166 | 2 |
| APRD00450 | Epinephrine | A01, B02, C01, R01, R03, S01 | 0.0216 | 0.472 | 158 | 3 |
| APRD00097 | Orphenadrine | M03, N04 | 0.0213 | 0.334 | 38 | 5 |
| APRD00267 | Tolbutamide | A10, V04 | 0.0190 | 0.269 | 27 | 2 |
| APRD01022 | Hydroxocobalamin | B03, V03 | 0.0190 | 0.325 | 16 | 8 |
| APRD00013 | Neomycin | A01, A07, B05, D06, J01, R02, S01, S02, S03 | 0.0189 | 0.488 | 187 | 2 |
| APRD00056 | Heparin | B01, C05, S01 | 0.0188 | 0.454 | 121 | 3 |
| APRD00174 | Clonidine | C02, N02, S01 | 0.0185 | 0.461 | 136 | 1 |
| APRD00326 | Vitamin B12 | A11, B03 | 0.0167 | 0.248 | 12 | 7 |
| APRD00536 | Vitamin B3 | C04, C10 | 0.0167 | 0.269 | 21 | 8 |
| APRD00650 | Procaine | C05, D04, J01, N01, S01 | 0.0164 | 0.478 | 199 | 5 |
| APRD00862 | Chloramphenicol | D06, D10, G01, J01, S01, S02, S03 | 0.0157 | 0.496 | 184 | 1 |

# Figure legends

**Figure 1:**

**a, b**: The therapy network at level 1 (a) and 2 (b). Nodes are colored according to the first level of the ATC classification. The size of nodes is proportional to the number of therapies in the class. The thickness of edges is proportional to the number of drugs linking the two therapies. **c**: Distribution of shortest path lengths in level 2 of the therapy network.

**Figure 2:**

Distribution of the number of ATC identifiers associated to each drug (corresponding to level 5 of the ATC classification.

**Figure 3:**

**a, b**: Degree distribution of the therapy network at level 2 (a) with degree exponent $\gamma = 0.76 \pm 0.10$ and level 3 (b) with $\gamma = 1.11 \pm 0.14$. **c, d**: Correlation between node degree $k$ and betweenness centrality $B_i$ (c) and closeness centrality (d) in the drug network at level 2. Correlation coefficient $r$ is indicated in figures. The P-value < 2.2e-16 in all cases.

**Figure 4:**

The top-20 drugs of highest betweenness centrality and their associated therapies at level 2. Drugs are represented by dark blue empty diamonds, therapies are represented by circles and colored following the same code as in figure 1.

# Additional files

### Additional file 1:

Full bipartite drug-therapy network corresponding to level 2 of the ATC classification. Drugs are represented by dark blue empty diamonds, therapies are represented by circles and colored following the same code as in figure 1 shown in main text.

### Additional file 2:

Full bipartite drug-therapy network corresponding to level 3 of the ATC classification.

### Additional file 3:

Full bipartite drug-therapy network corresponding to level 4 of the ATC classification.

### Additional file 4:

Distribution of shortest path lengths in level 3 of the therapy network.

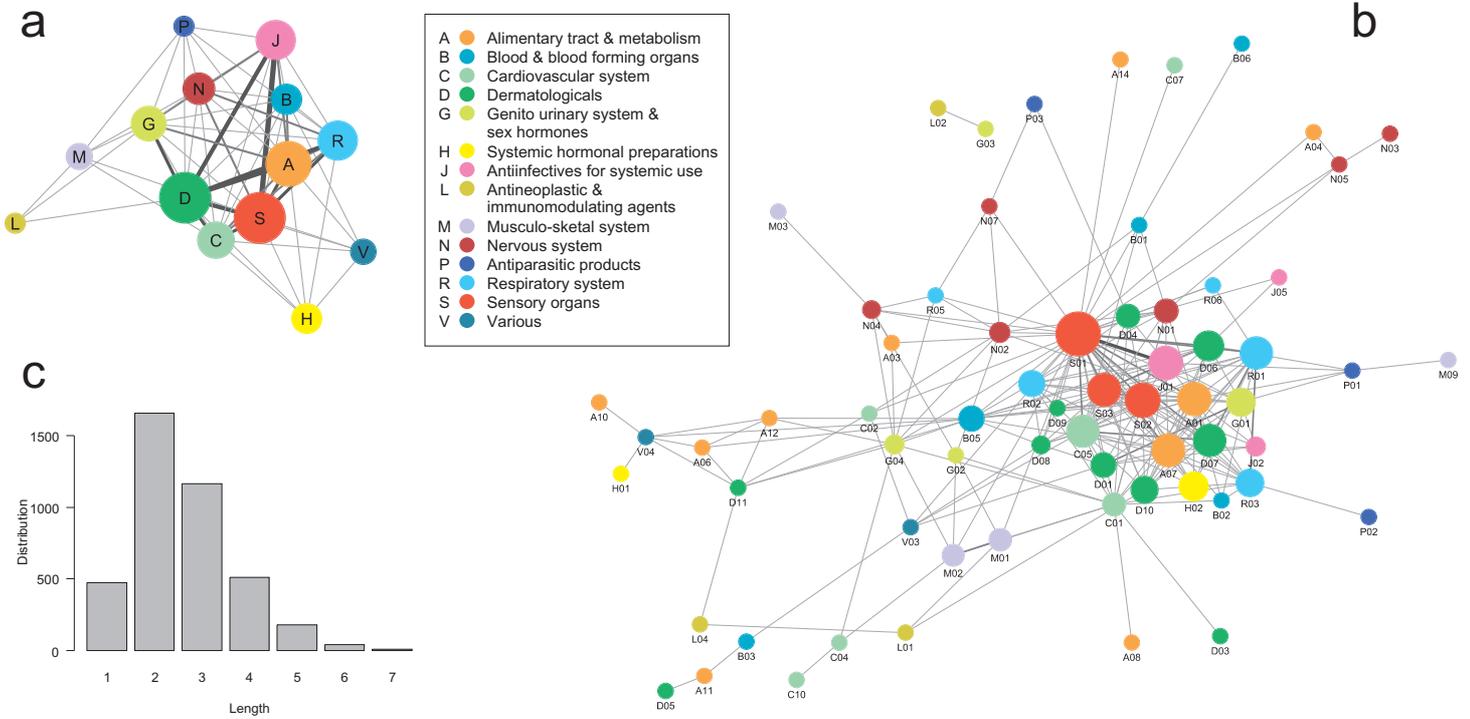

Figure 1

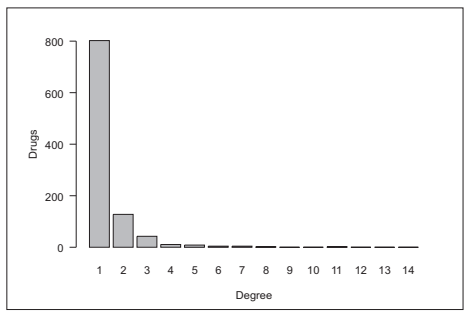

Figure 2

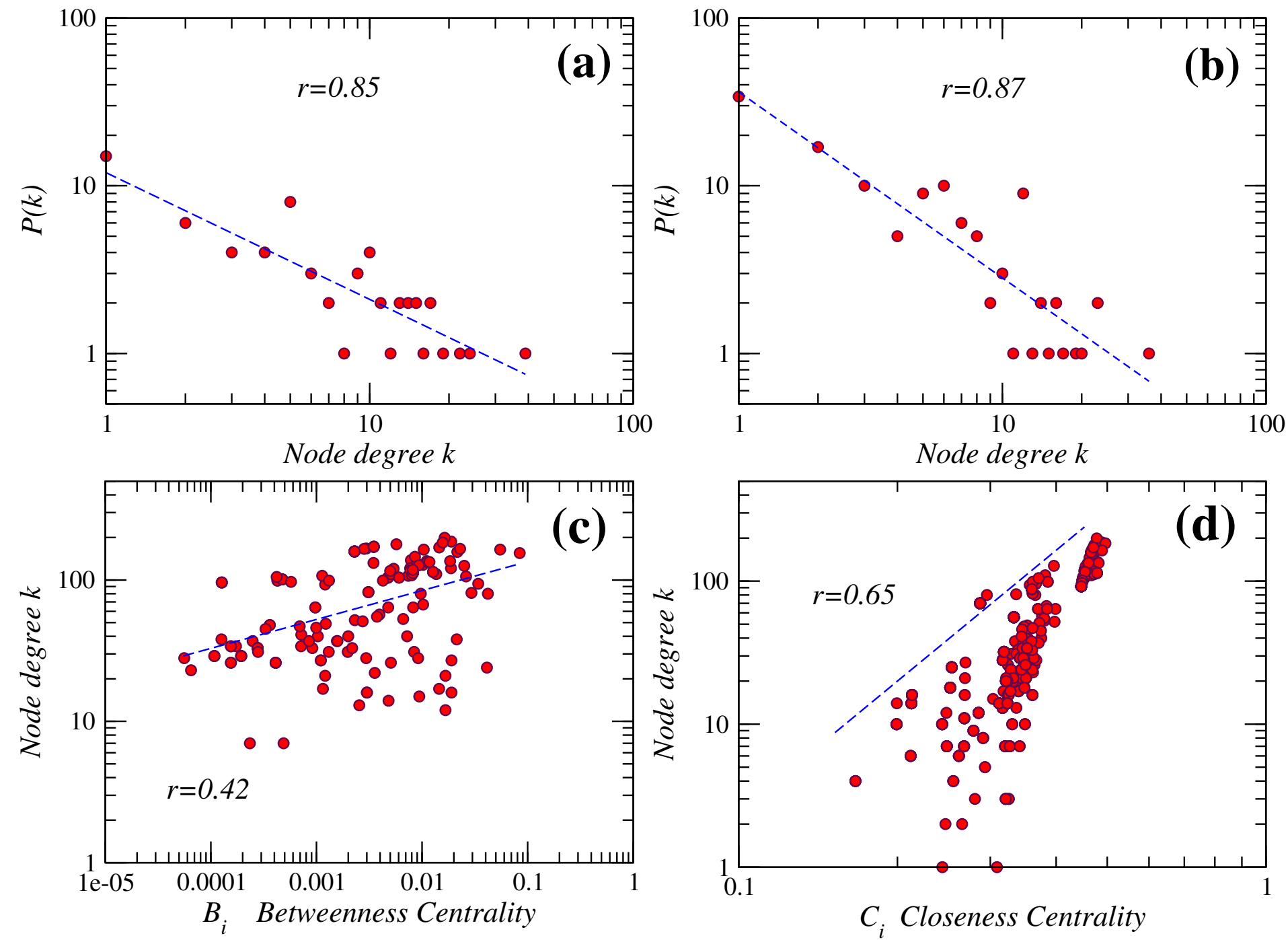

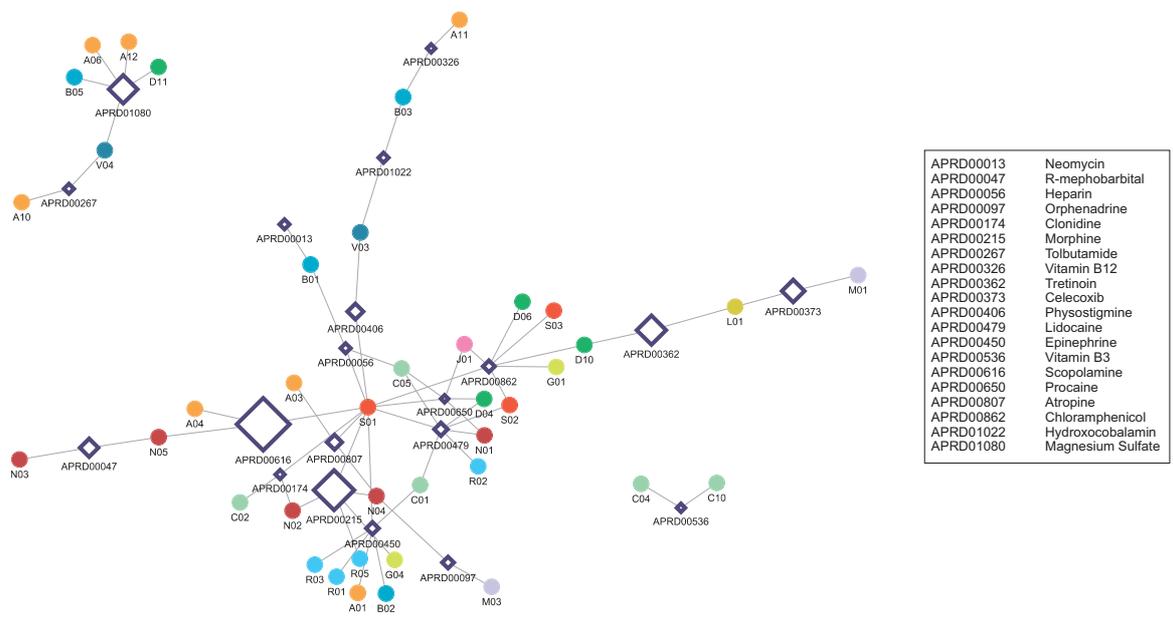

Figure 4

**Additional files provided with this submission:**

Additional file 1: bipartite2.pdf, 2533K
http://www.biomedcentral.com/imedia/3240518351593078/supp1.pdf
Additional file 2: bipartite3.pdf, 2671K
http://www.biomedcentral.com/imedia/2119445815930781/supp2.pdf
Additional file 3: bipartite4.pdf, 3048K
http://www.biomedcentral.com/imedia/1373485433159307/supp3.pdf
Additional file 4: figure-suppl4.pdf, 5K
http://www.biomedcentral.com/imedia/1000592681593078/supp4.pdf